\documentclass[a4paper,fleqn,usenatbib]{mnras}

\usepackage[T1]{fontenc}
\usepackage{ae,aecompl}

\usepackage{graphicx}	
\usepackage{amsmath}	
\usepackage{amssymb}	
\usepackage{newtxtext,newtxmath}

\title[A search for transits on \texttt{DIAmante} lightcurves]{A search for planetary transits on a set of 1.4 million multi-sector \texttt{DIAmante} lightcurves}
\author[M. Montalto]{
M. Montalto$^{1}$\thanks{E-mail: marco.montalto@inaf.it},
\\
$^{1}$INAF -- Osservatorio Astrofisico di Catania, 
Via Santa Sofia 78, Catania, 95123, Italy\\
}

\date{Accepted XXX. Received YYY; in original form ZZZ}

\pubyear{2017}

\begin{document}
\label{firstpage}
\pagerange{\pageref{firstpage}--\pageref{lastpage}}
\maketitle

\begin{abstract}
I report the results of a new search for transiting planets on a set of 1.4 million lightcurves extracted from {\it TESS} Full Frame Images (FFIs) using the \texttt{DIAmante} pipeline. The data come from the first two years of observations of {\it TESS} (Sectors 1-26) and the study is focused on a sample of FGKM dwarf and subgiant stars optimized for the search of transiting planets. 
The search was performed on the detrended and stitched multi-sector lightcurves applying the Box-fitting Least Squares algorithm and a Random Forest classifier.
I present a catalogue of 1160 transiting planet candidates, among which 842 are novel discoveries. 
The median radius of the transiting bodies in the catalog
is 6.8 R$_{\oplus}$. The radii range from 0.8 R$_{\oplus}$ to 27.3R$_{\oplus}$
while the orbital periods range from 0.19 days to 197.2 days with a median of 3.6 days.
Each candidate is accompained by a validation report and the corresponding DIAmante lightcurve. The material is available at CDS, on the ExoFOP website and on the DIAmante portal at MAST.

\end{abstract}




\section{Introduction}
\label{sec:introduction}

Following the launch of {\it TESS} \citep{ricker2014} I presented a new project to scrutinize the Full Frame Images (FFIs) downloaded from the satellite and to search for transiting planets with a new, dediacated pipeline named \texttt{DIAmante}. 
The results of the investigation of the first 13 sectors of data relative to the southern ecliptic hemisphere were discussed in a previous work \citep[][hereafter M20]{montalto2020}. 
Here, I present an update of this study, concerning the analysis of data coming from the first 26 sectors of {\it TESS} observations.



\section{Method} \label{sec:method}

\texttt{DIAmante} performs differential image analysis of {\it TESS} FFIs. The implementation I wrote involves a first order expansion of the differential background model which is fit simultaneously with the kernel coefficients to account for the variable background that often affects {\it TESS} images. On the subtracted images, I performed aperture photometry on four concentric circular apertures of 1,2,3 and 4 pixel radii respectively and centered on each target. The reference photometry to calculate the differential flux of each target was taken from the {\it Gaia} DR2 catalogue after calibrating it on the photometric reference system of {\it TESS}. 
Three sets of lightcurves were then generated for each target and each aperture photometry.
The LC0 lightcurves (which are the raw lightcurves), the LC1 lightcurves (corrected using a linear combination of eigenlightcurves) and the LC2 lightcurves (normalized by spline functions).

The detrending of each lightcurve against the set of eigenlightcurves is the same as in M20 as well as the stiching of all sectors where a given star was observed. 
Single sector observations represent 61 per cent of the sample, while lightcurves with two sectors represent 21 per cent, lightcurves with three sectors 7 per cent, and lightcurves with four or more than four sectors represent 11 per cent of the sample. Multiple sector observations are not necessarily contiguous and large gaps are present in the data.

On the LC2 lightcurves, the pipeline calculated the Box-fitting Least Squares  \citep[BLS,][]{kovacs2002} periodogram to search for transiting planets. 
The search was performed from a minimun frequency ($\rm f_{min}$) corresponding to the full time spanned by the lightcurves $\rm \Delta T$ ($\rm f_{min}=1/\Delta T$) to a maximum frequency ($\rm f_{max}$) obtained by setting the orbital distance equal to 2 $\rm R_{s}$, compatibly with the Kepler's third law:

\begin{equation}
    f_{max}=\sqrt{\frac{G M_{s}}{4 \pi^2 (2R_s)^3}},
\end{equation}

\noindent
where $G$ is the gravitational constant, $M_s$ the mass of the star, $R_s$ the radius of the star. I adopted the optimal frequency step reported in M20.

The identification of the transiting candidates was performed using a Random Forest classifier. 
As in M20, the classifier was applied on eleven features extracted from the lightcurves or from the periodograms: the signal-to-noise (S/N) of the primary eclipse (\texttt{snI}\footnote{
I note that the formula for the S/N of the primary eclipse was incorrectly typed in M20 since the term in parenthesis on the right hand side of Eq.~6 should have been at the denominator, i.e. 
$\rm SN_I=\frac{\delta_1}{\sigma}
\Big(\frac{1}{\sqrt{N_{in}}}+\frac{1}{\sqrt{N_{out}}}\Big)^{-1}$ and the same holds for Eq.~7, Eq.~8 and Eq.~11 of M20.
}), the S/N of the secondary eclipse (\texttt{snII}), the S/N of the tertiary eclipse (\texttt{snIII}), the signal detection efficiency 
(\texttt{sde}), the average signal detection efficiency of aliasing peaks (\texttt{sdeAL}),
the S/N of odd/even transits (\texttt{snOE}), the out-of-transit variability (\texttt{r2oot}), the fractional transit duration (\texttt{q}), the point-to-point statistics inside and outside the transit (\texttt{p2pio}), the simmetry of the folded lightcurve (\texttt{p2ps}) and the estimated transiting body radius (\texttt{r}).

The classifier was trained 
injecting simulated transiting planets as well as 
simulated eclipsing binaries and simulated short period sinusoidal variables on a representative set of {\it TESS} lightcurves. I did not trained the classifier to identify systematic effects in the dataset.
Like in M20, 
I decided to build a balanced training set constituted by
50 per cent of planets and 25 per cent of constant stars and the remaining 25 per cent equally subdivided between eclipsing binaries and short period sinusoidal variables. I used the maximum value of the Jouden's index \citep{youden1950} 
across the receiver operating characteristic (ROC) curve
to set the optimal threshold for the Random Forest classifier for each aperture. For aperture equal to 1 pix the threshold adopted implies a true positive rate (TPR) equal to 96.9$\%$ and a
false positive rate (FPR) equal to 2.3$\%$, 
for aperture equal to 2 pix TPR=97.7$\%$ and a FPR=3.2$\%$, 
for aperture equal to 3 pix TPR=97.7$\%$ and a FPR=3.7$\%$ and for aperture equal to 4 pix TPR=96.2$\%$ and a FPR=2.5$\%$. 
For apertures equal to 1 pix and 2 pix the thresholds adopted in this work implied a TPR definetly higher than in M20 (96.9$\%$ versus 92$\%$ for 1 pix aperture and 97.7\% versus 95.4$\%$ for 2 pix aperture in M20) and marginally higher false positive rates (2.3$\%$ for aperture 1 pix and 3.2$\%$ for aperture 2 pix, with respect to 1$\%$ in M20).
Different photometric apertures were used for stars corresponding to different {\it TESS} magnitudes ($T$). Aperture equal to 1 pix was used for $T\ge11$, aperture equal to 2 pix for 
$T<12.3$, aperture equal to 3 and 4 pix for $T<9.5$. This is a major difference with respect to M20 where only two apertures equal to 1 pix ( $T>11$) and 2 pix ($T\le$11) were used. If a transiting candidate was identified in  more than one aperture, I chose the one with the highest signal to noise of the primary transit. The reader is pointed to M20 for a more detailed description of the method I used.

\section{Stellar sample} 
\label{sec:sample}

The stellar sample is described in \citet[][hereafter M21]{montalto2021}. In particular the selection of FGK dwarfs and subgiants is reported in Appendix B of M21 while late K stars and M-dwarfs were selected imposing $M_{G,0}>3$, $G<15.5$ and $(G_{BP}-G_{RP})_0>1.65$
where $M_{G,0}$ is the absolute magnitude in the $G$ band corrected for reddening, 
$G$ is the apparent $G$ band {\it Gaia}-DR2 magnitude and  $(G_{BP}-G_{RP})_0$ is
the {\it Gaia}-DR2 colour corrected for reddening.
The stellar parameters of FGK dwarfs and subgiants were taken from the all-sky PLATO Input Catalogue (asPIC1.1, M21) when present in it or derived with the same method, while for M-dwarfs I took stellar parameters from the {\it TESS} Input Catalogue \citep{stassun2019}. In total, I analysed 1 407 969 stars. None of the objects analysed in this work was in common with M20.

\section{Vetting} 
\label{sec:vetting}

\begin{figure}
\includegraphics[width=\columnwidth]{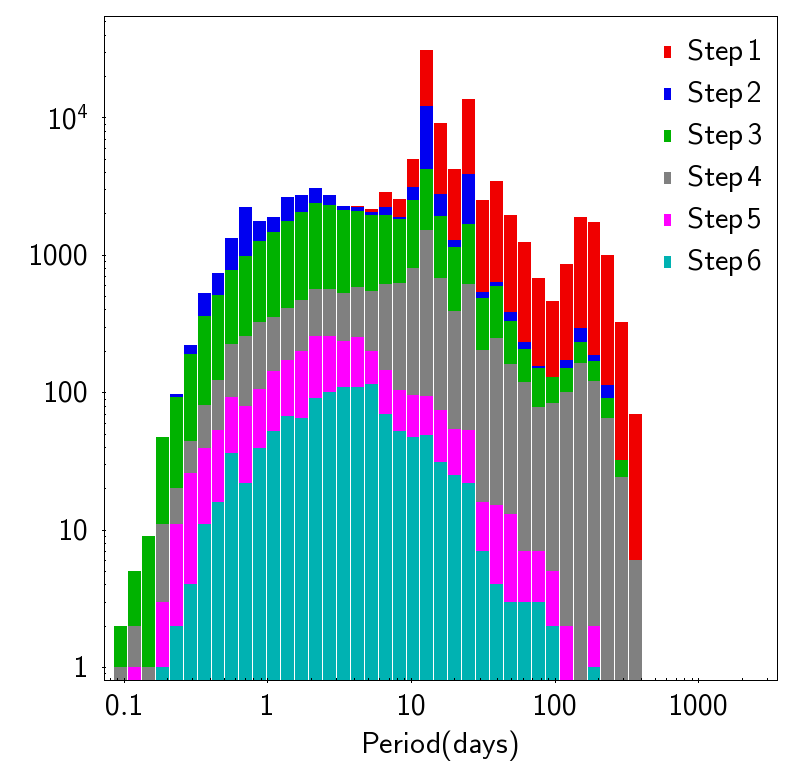}
\caption{
Orbital period distribution of candidates selected during the different vetting steps (see text).
}
\label{fig:vettingsteps}
\end{figure}

I followed a multi-step procedure to vet the list of candidates. Figure~\ref{fig:vettingsteps} shows the period distribution of the candidates corresponding to each step of the procedure as explained in the following text and denoted by the different colours of the histograms.
The Random Forest classifier thresholds were passed by 110712 stars (step 1).

By inspecting the results, I found that a large number of candidates were associated with events with less than three transits having long periods and that most of them were spurious artefacts so I decided to retain candidates with two or less transits only if they had a random forest probability $>70\%$, which reduced the list to 56576 stars (step 2).

Then I checked for the presence of exceedingly large number of candidates at any given period, which is usually a signature of possible systematics present  in the data. I considered groups of 10000 randomly chosen stars and build the distribution of th candidates' BLS periods adopting 0.1 days period bins between the minimum and maximum detected period. I then retained only those candidates belonging to period bins where the number of detected objects was $<$Q$_3+1.5\times$IQR where Q$_3$ was the third quartile of the distribution of the number of targets detected in each period bin and IQR the interquartile range. After this selection 38075 stars remained (step 3). The most prominent systematics appeared at periods corresponding to the duration of a {\it TESS} sector ($\sim$27 days) or half of it ($\sim$13.5 days).

 I then run the same centroid motion algorithm described in M20 where two probabilities were derived: P$\rm_{D}$, the probability that the source of variability was associated with the target given the observed centroid motion measurements distributions and $P_{\eta}$, the reliability of the centroid motion results considering the local density of stars surrounding the target. The algorithm calculates also the closest star to the centroid position in each aperture.
As in M20, I imposed $P\rm_{D}>50\%$ and $P_{\eta}>30\%$ and that in at least three out of four apertures the closest star to the centroid position was the target itself. After this selection 11713 objects remained (step 4). The centroid motion algorithm eliminated 93\% of the candidates with Galactic latitude $-5^{\circ}<$b$<5^{\circ}$ representing about 30$\%$ (11051) of the stars selected in step 3. This is due to the fact that the adopted detection thresholds permit to avoid the selection of stars in the crowdest regions of the sky favouring the selection of targets that are less affected by contamination from neighbouring stars.

I proceeded by a visual inspection of each one of the targets, eliminating those that were most obvious false positives and in particular variables with rather complex variability patterns that were not properly normalized or candidates where outliers from the reduction procedure were by chance concidence aligned by the folding algorithm generating spurious transits.
The remaining list I obtained after this step contained 2813 stars (step 5).

In the next step, I quantified the transit depth difference on adjacent apertures. The \texttt{apstat} parameter was defined as

\begin{equation}
    apstat=\frac{|\delta_A-\delta_B|}{\sqrt{\sigma^2_A+\sigma^2_B}}
\end{equation}

\noindent
where $\delta_A$, $\delta_B$ are the transit depths measured on the lightcurves relative to two adjacent apertures (usually the best aperture and the closest larger aperture) and $\sigma_A$, $\sigma_B$ are the standard deviations of the respective out-of-transit measurements.
The parameter \texttt{apstat} was imposed to be \texttt{apstat}$<5$.

Additionally, I also retrieved the value of the renormalised unit weighted error ($\texttt{ruwe}$) from the {\it Gaia} DR3 \citep{gaiadr32022} archive and impose it to be smaller than \texttt{ruwe}$\leq1.4$. Following the {\it Gaia} DR3 documentation, I then also checked for the presence of radial velocity variable stars in the sample. These objects were identified imposing the radial velocity P-value for constancy (\texttt{rvChiSqPValue})$<0.01$, the radial velocity renormalised goodness of fit (\texttt{rvRenormalisedGof})$>4$ and the number of transits used to compute the radial velocity (\texttt{nrv})$\geq10$. 
Stars with poor astrometric solutions were identified as in M20, where confirmed binaries were found imposing 
the Significance of the Astrometric Excess Noise (\texttt{astroExcessSig}) $>$5 and the 
Goodness of Fit in the Along-Scan direction (\texttt{gofAl})
$>$20.
Potential radial velocity variables or stars with 
poor astrometric solutions were flagged in the catalog (\texttt{binflag}$=$1) not eliminated, and are 14$\%$ (160 stars) of the total number of stars.  
Then I checked for the consistency of the odd and even transits (\texttt{snOEfit}$<5$), for the absence of secondary eclipses (\texttt{snII}$<5$) and checked if any object was a known false positive (see below). After imposing these conditions 1160 objects remained (step 6).
For all these candidates, I performed a Levemberg-Marquardt fit with a \citet{mandel2002} model and with bootstrap resampling to estimate the physical properties of the candidates and their uncertainties.

\begin{figure*}
\includegraphics[width=15cm]{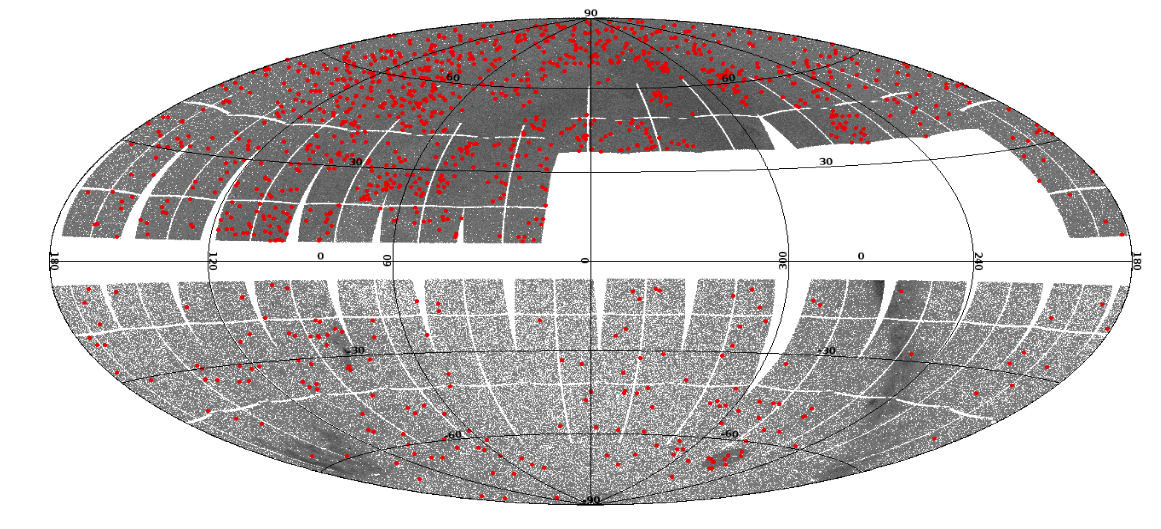}
\caption{
Spatial distribution across the celestial sphere of all sources analyzed in this work (grey points) and of the transiting planet candidates (red dots) in an ecliptic coordinates reference system. 
}
\label{fig:spatial_distribution}
\end{figure*}

\begin{figure}
\includegraphics[width=7.5cm]{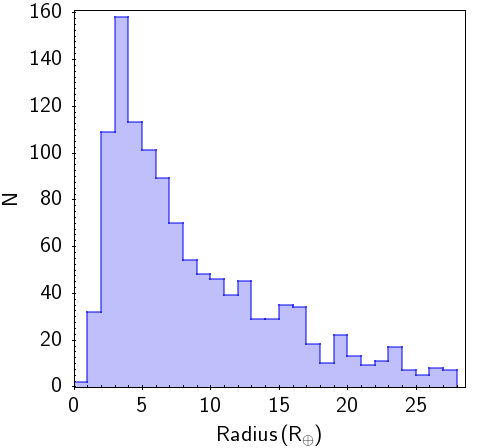}
\caption{
Radius distribution of the transiting planetary candidates. 
}
\label{fig:period_distributions}
\end{figure}

\begin{figure}
\includegraphics[width=7.7cm]{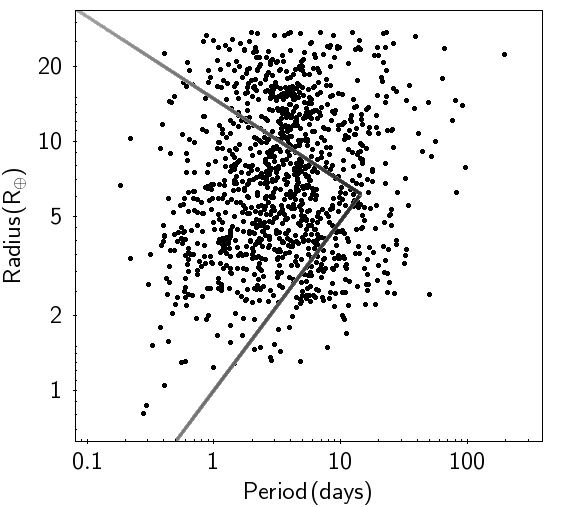}
\caption{
Radius versus orbital period diagram for the transiting planetary candidates discussed in this work. The grey lines denote the boundaries of the Neptune desert defined in \citet{mazeh2016}.
}
\label{fig:radius_period}
\end{figure}

\section{Comparisons} 
\label{sec:comparisons}

I compared the list of candidates reported in this work with other catalogs of confirmed planets, candidates planets or false positives publically available.  I retrieved from the NASA Exoplanet archive the list of 3902 confirmed transiting exoplanets (accessed August 14, 2022). Of these 365 are included in the starting (1.4 million stars) catalog and 61 are present in the final catalog, yielding a 17\% confirmed transiting planets recovery rate. 
The cross-match with the current (accessed August 21, 2022) list of 5845 TOIs resulted in 304 common entries out of the 1261 TOIs present in the 
starting catalog\footnote{\url{https://tev.mit.edu/data/collection/193/}}
, yielding a TOI recovery rate equal to 24$\%$, while the cross-match with the list of 1926 cTOIs (accessed August 21, 2022) resulted in 46 commmon entries out of 468 present in the starting catalog (10$\%$ recovery rate)\footnote{\url{https://exofop.ipac.caltech.edu/tess/view_ctoi.php}}.
Recently, \citet{olmschenk2021} presented a list of transiting planetary candidates found in {\it TESS} Full-Frame images using convolutional neural networks. The starting catalog contained 39 of these candidates and 15 were recovered and are in the candidate's catalog yielding a 38$\%$ recovery rate. \citet{eisner2021} presented the results from the first two years of the Planet Hunters {\it TESS} citizen science project reporting 90 most promising candidates. Of these, 47 are present in the starting catalog, and 3 were recovered (6$\%$ recovery rate). The three objects recovered are 
 229742722 for which I found a period of $\sim$63.5 days against the 29 days reported by the authors, 328933398.01 and 274599700. \citet{yu2019} modified an existing neural network designed for {\it Kepler} and applied it to {\it TESS} data. Among their 288 candidates 15 are present in the initial catalog and one was recovered, star 421900585 . The cross-match with the catalog of {\it K2} targets observed by \citet{dressing2019} renstituted one match with the initial catalog, star with EPIC number 212204403 which was however not recovered in the final catalog.
 Similarly, the cross-match with the catalog of bright stars in {\it K2} campaigns 0-10 of \citet{mayo2018}, yielded two matches, EPIC 211418290 and EPIC 228729473, none of which was recovered. 

 The cross-match of the final Q1-Q17 DR25 {\it Kepler} KOI table\footnote{
 \url{https://exoplanetarchive.ipac.caltech.edu/cgi-bin/TblView/nph-tblView?app=ExoTbls&config=q1_q17_dr25_koi}
 } \citep{thompson2018} with the starting catalog includes 655 objects, among which three are then present in the candidates' catalog: TIC 275489982 (KOI 6066.01) and TIC 26581967 (KOI 1487.01) which are known false positives and TIC 417676622 (KOI 246.01) which is a confirmed planet ({\it Kepler}-68b) for which I confirmed both the period and the radius. 
 
 The cross-match of the initial catalog with the list of binaries provided by \citet{twicken2016} in the {\it Kepler} field, and excluded from the transit search of {\it Kepler}, includes 67 stars, none of which was  present in the final catalog. The centroid crowding algorithm strongly reduced the number of candidates in low Galactic latitude fields like the one of {\it Kepler}. I also checked the catalogs of \citet{vonBoetticher2019,feinstein2019,dong2021}, but none of the objects reported in these studies were included in the initial catalog discussed here.  
 
I compared the list of candidates with the KELT false positives catalog presented by \citet{collins2018}.  The search resulted in four matches: 44631965 (SB1), 115618085 (EB1), 55655482 (SB1), 100009973 (SB2). The total number of KELT false positives present in the starting catalog is equal to 298 stars which gives a 1.3$\%$ false positives recovery rate. I did a similar search using the SuperWASP dispositions and False Positive Catalogue \citep{schanche2019} which resulted in five matches (among which one is in common with the KELT list of false positives): 44631965 (EBLM), 9828416 (blend), 193628179 (EB), 157850336 (EBLM),
270515566 (EBLM). The number of SuperWASP false positives present in the starting catalog is 279 which yields a 1.8$\%$ false positives recovery rate.

I then searched for possible contaminating eclipsing binaries and variables blended with the candidates using data from the Zwicky Transitnt Facility \citep[ZTF, ][]{Bellm2018}. I downloadad from the ZTF website a catalog containing all sources having r-band magnitude $<20$, within a radius of one {\it TESS} pixel (21 arcsec) from each candidate, having an associated $\chi$ square (of the corresponding ZTF lightcurve) $>1.5$ and with at least 50 measurements. I then retrieved the corresponding ZTF lightcurves. In total, I retrieved 2206 lightcurves associated to 589 candidates. For each lightcurve, I run the BLS algorithm fixing the period at the candidate's period, half or twice that period. For candidates with only one or two transits I searched for periods within 0.25 and 50 days. I then flagged those objects for which the S/N calculated by the BLS was larger than 6. This resulted in 70 possible contaminants. I then visually inspected the lightcurves of all these stars and compared them with the lightcurves of the candidates to which they were associated concluding that 13 where likely at the origin of the observed signal. The remaining stars were either ambiguos signals not directly related to the target's signal or spurious artefacts. The 13 candidates associated with the identified contaminating sources were the stars with TIC identification number:
240009525, 28643432, 100633294, 115473018, 174094041, 252912175, 375649193, 372071520, 63008393, 9484064, 154702446, 265887000, 8408706. 
All the false positives I identified have been removed from the final catalog.

Despite having passed several vetting steps, the objects I present should still be considered candidate planets because their masses are unknown, because they can be the results of unresolved blends or binary/multiple systems \citep[e.g.][]{morton2011,santerne2013} or even unrecognized systematics. Since a large fraction of the candidates' hosts are relatively bright (484 targets have {\it V}<12)
spectroscopic follow-up and high resolution imaging can be pursued to better investigate the nature of the candidates.

\section{Results} 
\label{sec:results}

The search produced 1160 candidates among which 842 are novel discoveries. The median radius of the transiting bodies in the catalog is 6.8 R$_{\oplus}$, a significant improvement with respect to M20 where the median radius was 11.5  R$_{\oplus}$. The radii range from 0.8 R$_{\oplus}$ to 27.3R$_{\oplus}$ while the orbital periods range from 0.19 day to 197.2 days with a median of 3.6 days. 
Fig.~\ref{fig:spatial_distribution} presents the spatial distribution of all stars analyzed in this work (grey points) and of all candidates (red dots) in an eclipctic coordinates reference system. Fig.\ref{fig:period_distributions} presents the radii distribution of the transiting planets candidates while Fig.~\ref{fig:radius_period} shows the radius versus period diagram.


The catalog contains 81 candidates around bright stars ({\it T}$<$9). Among these 55 are new detections. 
The candidate found around the brightest star (332064670, {\it T}=5.2) is the well known 55 Cnc e \citep{winn2011}.


Ultra-short period \citep[USP,][]{sanchis2014} planets
are planets with orbital periods smaller than one day.
Around solar type stars they are thought to be as frequent as Hot-Jupiters and to have radii smaller than 2 R$_{\oplus}$ \citep{winn2018}. The catalog contains 11 objects with period$<$1 day and radius R$<$2 R$_{\oplus}$, among which 6 are new detections. However, there are overall 131 USP candidate planets in the catalog with period $<$1 day (56 with radius R$<$4 $\rm R_{\oplus}$) among which 120 are new detections and 25 USP candidates are found around late type K or M stars (T$\rm_{eff}$<4000 K). The candidate with the smallest radius in the catalog has an orbital period equal to 0.27 days and a sub-Earth size radius R$\sim$0.8 $\rm R_{\oplus}$ and it is orbiting a relatively bright (372207328, {\it T}=10.97) very late type M-dwarf stars.

Another interesting population of planets is that one of Hot-Neptunes.
\citet{szabo2011} noted the lack of Hot-Neptunes with periods P$<$2.5 days in the population of planets known at the time. By adopting the equations reported by \citet{mazeh2016} which define the boundaries of the Neptune desert in the period-radius diagram (see Fig.\ref{fig:radius_period}), I found that 612 of the TESS candidates (516 new detections) reported in this work lie in the desert. 

Moreover, there are 280 candidates (253 new) found around subgiant stars with stellar radius R$>$2 R$_{\odot}$, 23 (12 new) long-period (P$>$30 days) candidates and nine (seven new) single transit candidates.




\section{Conclusions} 
\label{sec:conclusions}

I presented a new search for transiting planets 
on a set of 1.4 million optimally selected FGKM dwarf and subgiant stars observed during the first two years of the TESS mission (Sector 1-26). 
The search produced 1160 transiting planetary candidates among which 842 are new detections.

The method I used to identify the transiting planet candidates is similar to the one presented in M20 with improvements related to the adoption of more optimized detection thresholds and the use of large apertures for bright stars. Overall, this led to the detection of smaller size objects than in M20. 

Other improvements can be made to the pipeline. The Random Forest classifier could be applied to a larger pool of features to obtain higher sensitivity and lower false positive rate.
The centroid motion algorithm could be revised considering less restrictive conditions to permit the analysis of the Galactic Plane.
Autocorrelation analysis of the residuals after normalization of the lightcurves could be used to identify repetitive patterns still lingering in the data \citep[e.g.][]{caceres2019} which could be modeled and removed before performing the transit search.

\section{Acknowledgements}

This paper includes data collected by the TESS mission, which are publicly available from the Mikulski Archive for Space Telescopes (MAST). Funding for the TESS mission is provided by NASA’s Science Mission directorate. This research has made use of the Exoplanet Follow-up Observation Program website, which is operated by the California Institute of Technology, under contract with the National Aeronautics and Space Administration under the Exoplanet Exploration Program.

\section{Data availability} 
\label{sec:catalogue}

I released the catalog of all candidates found in this work at CDS. The lightcurves of the newly discovered candidates along with their validation reports are submitted to the ExoFOP portal at \url{https://exofop.ipac.caltech.edu/tess/}. The catalog of all candidates, all validation reports and lightcurves can be also found on the webpage dedicated to the \texttt{DIAmante} project at MAST at \url{doi:10.17909/t9-p7k6-4b32} and  \url{https://archive.stsci.edu/hlsp/diamante}.




\bibliographystyle{mnras}
\typeout{}
\bibliography{example} 

\bsp	
\label{lastpage}
\end{document}